# Integration of RTMFP in the OMNeT++ Simulation Environment

Felix Weinrank
Münster University of Applied Sciences
Dept. of Electrical Engineering
and Computer Science
Bismarckstrasse 11
D-48565 Steinfurt, Germany
weinrank@fh-muenster.de

Michael Tüxen
Münster University of Applied Sciences
Dept. of Electrical Engineering
and Computer Science
Bismarckstrasse 11
D-48565 Steinfurt, Germany
tuexen@fh-muenster.de

Erwin P. Rathgeb
University of Duisburg-Essen
Institute for Experimental Mathematics
Ellernstrasse 29
D-45326 Essen, Germany
erwin.rathgeb@iem.uni-due.de

*Abstract*—This paper introduces the new Real-Time Media Flow Protocol (RTMFP) simulation model for the INET framework for the OMNeT++ simulation environment. RTMFP is a message orientated protocol with a focus on real time peer-to-peer communication. After Adobe Inc. released the specifications, we were able to implement the protocol in INET and compare its performance to the similar Stream Control Transmission Protocol (SCTP) with a focus on congestion control and flow control mechanisms.

## I. INTRODUCTION

Real time audio and video communication has become a more and more important topic over time. Especially the increasing market penetration of mobile devices like smartphones and tablets presents new challenges for IP based communication protocols concerning address changes due to network handover, encryption and low latency. Protocols like the Real-Time Media Flow Protocol (RTMFP) [1] and WebRTC [2] have been developed to face these requirements. As Adobe published the specifications for RTMFP it was possible to implement an RTMFP model in the OMNeT++ / INET suite.

This paper will introduce RTMFP, its integration into the INET framework, the model validation and a comparison with the similar Stream Control Transmission Protocol (SCTP) [4].

Section II of this paper will give an overview of RTMFP itself, followed by the description of the protocol integration in the INET framework. Section IV will cover the validation of the new model followed by a comparison of RTMFP and SCTP with regard to performance and fairness.

## II. THE REAL-TIME MEDIA FLOW PROTOCOL (RTMFP)

The Real-Time Media Flow Protocol (RTMFP), specified in RFC 7016 [1], is a peer-to-peer protocol for real time communication over IP based networks and has been developed by Adobe Systems Inc.

RTMFP is widely deployed on several platforms due to its integration in the Adobe Flash Player and Adobe Air, where it is implemented since 2008, whereas SCTP is an essential part of WebRTC where it is used for the WebRTC

Data Channels [5]. The RFC 7016 did not specify specific encryption methods, these have been supplemented by Adobe in the RFC 7425 [3].

Its use cases are audio and video calls with low latency as well as bulk data transfers. In the beginning, the specifications of RTMFP were closed source but after several drafts, Adobe published the final RFC 7016 in November 2013. The message oriented, UDP based protocol contains features for NAT traversal and is encrypted by default - only address and session information remain unencrypted.

### A. Sessions, Flows and Messages

A connection between two RTMFP peers is called session. Each session is bidirectional, congestion controlled and identified by its unique session identifier. In order to meet the demands of mobile environments, RTMFP supports an IP address change of a peer and connection re-establishment without interruption. The usage of a unique session identifier allows an address change of one peer without re-establishing the connection - the sender just continues sending packets and the receiver will recognize the peer address change and use the new address. To resist denial-of-service attacks, a four-way-handshake is used.

Sessions can contain zero or more unidirectional, message oriented flows for data transmission which are identified by a unique flow identifier. Flows are multiplexed, prioritized, allow in- or out-of-order delivery with variable reliability, and every flow has an independent flow control.

Each RTMFP message consists of a generic header and one or more chunks. The bundling of chunks reduces the protocol overhead especially if many small fragments are transmitted.

### B. Congestion- and Flow-Control

Since RTMFP is UDP based, it has to implement its own congestion control and flow control. The built-in loss based and TCP-friendly congestion control manages an independent congestion window for each session and allows the prioritization of time critical data like audio or video over bulk data. When sending real time data on one session, the congestion control informs other local sessions and its peer about the time critical transfer. As a consequence, the other congestion

Parts of this work have been funded by the German Research Foundation (Deutsche Forschungsgemeinschaft).

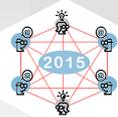

controls should switch to a less aggressive mode to prioritize real time data.

The congestion control is driven by acknowledgements and packet loss indications, sent by the receiver, and by timeouts. When the receiver acknowledges received packets to the sender, the sender increases its congestion window. If the sender detects packet loss due to loss indications by the receiver or timeouts, it will reduce the congestion window in relation to its operation mode. When the session operates in real time mode, the congestion control increases the window in a more aggressive way than in normal operation mode. The same procedure is applied for reacting on packet loss where the congestion control will reduce the congestion window in a less aggressive way in real time mode than in normal mode.

Every flow has its own flow control to manage the receiver buffer. In every acknowledgement chunk the receiver announces the available receive buffer size to the sender.

### C. RTMFP and SCTP

RTMFP is quite similar to SCTP in many aspects. Both protocols use a four-way-handshake, support bundling and fragmentation and have the concept of streams/associations and flows/streams. Also like RTMFP, SCTP has an independent congestion control per association. In contrast to RTMFP, SCTP does not have built-in encryption but this can be achieved by the use of Datagram Transport Layer Security (DTLS) [6] like in the WebRTC stack where SCTP packets are encapsulated by DTLS.

### III. THE SIMULATION MODEL

The RTMFP simulation model for the INET framework consists of two simple modules called RTMFP-Layer and RTMFP-App, together they build the RTMFP-Compound-Module. An RTMFP-Compound-Module consists of one RTMFP-Layer instance and one or more RTMFP-Apps and is connected to the INET Standardhost by a UDP gate - see Figure 1. As we were particularly interested in the transmission performance of the protocol, we focused on the connection establishment, chunk bundling, flow control and the congestion control. These features are completely covered by the simulation model, wheras features like NAT traversal or encryption are not implemented yet.

### A. RTMFP-App

The RTMFP-App module simulates an application which makes use of RTMFP. Each RTMFP-App represents an independent endpoint that can receive data and may also act as a sender to initiate a session to another endpoint with one or more independent flows. Each flow is configured separately with respect to its packet rate, send interval, maximum number of packets and its absolute prioritization. The RTMFP-App records statistical data for every flow like transfer rate, number of received and sent messages and runtime.

### B. RTMFP-Layer

The RTMFP-Layer represents the operation of RTMFP. All RTMFP specific features like connection establishment, session management, flow control and congestion control have been implemented in this module. The strict separation in RTMFP-Applications and RTMFP-Layer allows us to quickly create new applications for testing scenarios while keeping them as simple as possible.

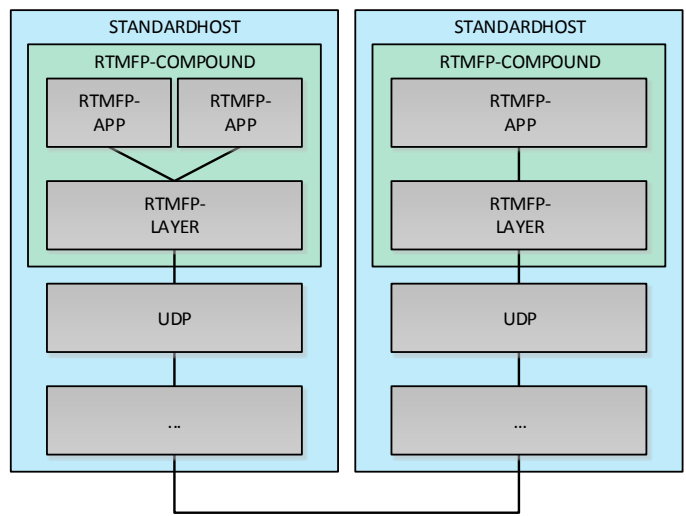

Fig. 1. RTMFP-Compound-Module architecture

### C. Model Parameters

To cover a wide range of peer-to-peer communication scenarios from single file transfer to more complex scenarios like audio and video conferences with parallel file transfer, it is necessary to make the configuration as customizable as possible. We fulfilled this requirement by a flexible architecture where the user can add any number of RTMFP-Apps with any number of flows by declaring them in the configuration file.

Figure 2 shows a sample configuration for an RTMFP-Compound-Module. `numRtmfpApps` defines the number of RTMFP-Apps attached to the RTMFP-Layer.

The RTMFP-Layer can be configured by setting its UDP port (`localPort`), maximum packet size (`maxSegmentSize`), the initial congestion window (`ccCwndInit`) and the size of the receive buffer for incoming flows (`rcvBufferSize`). The initial congestion window is set by `ccCwndInit`.

Every RTMFP-App has its own subset of settings: `flowsOutgoing` sets the number of outgoing flows. All outgoing flows are configured by a list of parameters, separated by spaces. The size of every outgoing packet (`flowPacketSize`), the interval between every send call (`flowSendInterval`) and the amount of outgoing packets (`flowNumPackets`) can be configured as fixed values or by using any OMNeT++ built-in distribution like `exponential` or `uniform`. If a built-in distribution is chosen it will be evaluated by every call of the RTMFP-App send function. `readDelay` is the timespan the application will wait to request data from the layer after being notified about arrived data, this is useful to fill the receive buffer.

```
# HOST 1
**.udpApp[0].numRtmfpApps = 2
**.udpApp[0].layer.localPort = 4711
**.udpApp[0].layer.maxSegmentSize = 1472byte
**.udpApp[0].layer.rcvBufferSize = 65536byte
**.udpApp[0].layer.ccCwndInit   = 4380byte

# APP 0
**.udpApp[0].app[0].localEpd = 4712
**.udpApp[0].app[0].remoteAddress = "host2"
**.udpApp[0].app[0].remotePort = 2013
**.udpApp[0].app[0].remoteEpd = 2014

**.udpApp[0].app[0].flowsOutgoing = 2
**.udpApp[0].app[0].flowPacketSize = "140byte 140byte"
**.udpApp[0].app[0].flowSendInterval = "1000us 1000us"
**.udpApp[0].app[0].flowNumPackets = "500000 500000"
**.udpApp[0].app[0].flowTimeCritical = "1 1"
**.udpApp[0].app[0].flowId = "19 88"
**.udpApp[0].app[0].maxRuntime = 1800s
**.udpApp[0].app[0].readDelay = 0ms

# APP 1
...
```

Fig. 2.  Simulation parameters for an RTMFP-Layer and an RTMFP-App with two outgoing flows

### D. Model Operation

When starting the simulation, the RTMFP-Apps register themselves at the layer with their unique Endpoint-Discriminator (EPD) and, if configured to do so, they trigger the RTMFP-Layer to establish a session with the given peer, also identified by its EPD, and one or more address candidates. When the RTMFP-Layer has successfully established a connection, it informs the RTMFP-App about the new session and the RTMFP-App starts sending data to the peer. The RTMFP-Layer and RTMFP-Apps use a custom set of cMessages to communicate. When an RTMFP-App is configured to start a data transfer to a peer, it starts sending data messages to the RTMFP-Layer which are queued in the given send flow.

If a single RTMFP-App message exceeds the Maximum Transmission Unit (MTU) it will be fragmented by the sending RTMFP-Layer into matching chunks and reassembled by the receiving RTMFP-Layer. Despite other attributes, every queued message gets an increasing sequence number, an indication if it is a complete message or a fragment and a priority marker, i.e. time critical or not.

If the flow control of the specific flow and the congestion control allow sending data, the RTMFP-Layer tries to bundle as many chunks as possible into one RTMFP packet - preserving the MTU.

When receiving a data chunk from a peer, the RTMFP-Layer will put the payload into the receiver queue of the corresponding receive flow and send a notification to the associated local RTMFP-App. The receiver should acknowledge every second received data message by an acknowledgment chunk, unless a packet loss has been detected.

When receiving an acknowledgment chunk, the sender deletes successfully transmitted messages from its queue and adapts the flow control and the congestion window. A message that is reported missing three times should be handled as lost and be retransmitted by the sender. This will all be done by the RTMFP-Layer, the RTMFP-App will not be notified. The acknowledgment chunks have a direct impact on the congestion control, every time the receiver reports a successful transmission without a loss report, the congestion control increases the congestion window, depending on its operation mode. In the case that a packet loss is detected by a loss report, the congestion control decreases the congestion window.

After all messages have been sent or the maximum simulation runtime is reached, all RTMFP-Layer and RTMFP-Apps record their statistics and shut down.

## IV. MODEL VALIDATION

To validate the RTMFP protocol implementation, we examined a set of predictable simulation scenarios to test the implemented features - especially flow control, congestion control and bundling were in our focus.

Due to the lack of comparable RTMFP models and the fact that RTMFP is only available as a commercial product via Adobe Flash or Adobe Air, it was not possible to cross-validate our model with real implementations.

To achieve a more realistic testing scenario and get some randomness, we used a bottleneck link scenario where the RTMFP traffic competed with a random UDP sender on the bottleneck which sent a small amount of traffic - see Figure 3.

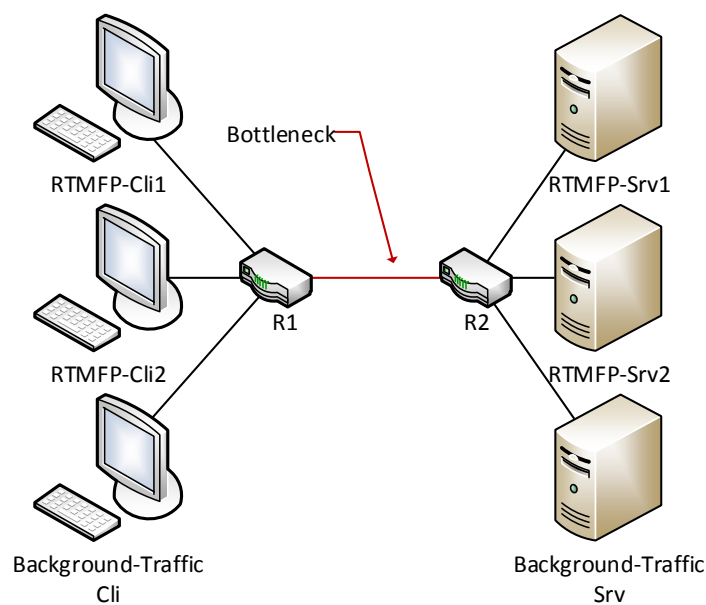

Fig. 3.  Bottleneck scenario with two RTMFP peers and background traffic

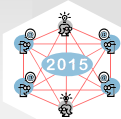

- The **flow control** has been validated by using the bandwidth delay product (BDP) which sets the bandwidth in relation to the link delay. We used a bottleneck scenario with a variable delay from 0 to 100 milliseconds, measured the bandwidth and compared the results with our theoretical calculations.

- Two or more competing RTMFP-Clients share a bottleneck for **congestion control** testing. In the first scenario all clients start sending simultaneously. See Figure 4. In the second scenario the clients start sending successively. The congestion control mechanisms should ensure a fair distribution of the available bandwidth.

- We varied the size of the applications data messages and recorded the bandwidth to prove the **bundling** mechanism. As every data chunk has a given header size, small data chunks have a low payload to header size ratio and this will result in a lower payload transfer rate.

The testing scenarios reached our expectations.

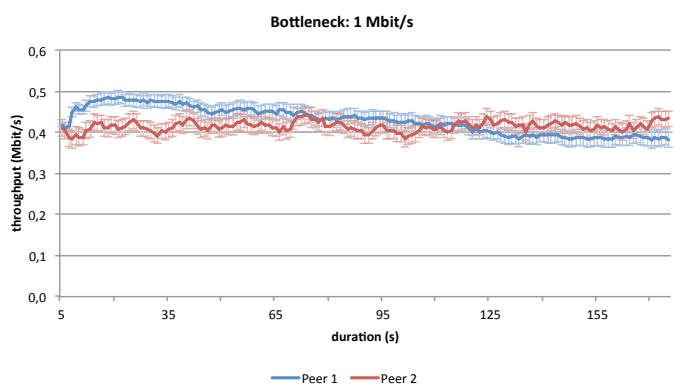

Fig. 4. Bottleneck scenario with two RTMFP peers and random traffic

## V. Comparing RTMFP to SCTP

After we had tested the RTMFP model in relevant aspects, we were able to compare RTMFP with the OMNeT++ / INET SCTP model [7]. One of our primary interests was the comparison of RTMFP and SCTP in relation to fairness and performance under different conditions like bottleneck-scenarios, lossy and delayed links.

Figure 5 shows the impact of packet loss on the transfer rate of both protocols. Whereas both protocols show a similar bandwidth performance in a scenario without loss, SCTP delivers nearly double the amount of data under packet loss conditions.

In direct comparison with a bottleneck scenario, RTMFP and SCTP share the available bandwidth nearly fair under most scenarios. When RTMFP operates in the normal operation mode - without real time traffic - the congestion control is less aggressive than the SCTP congestion control, which meets our expectations.

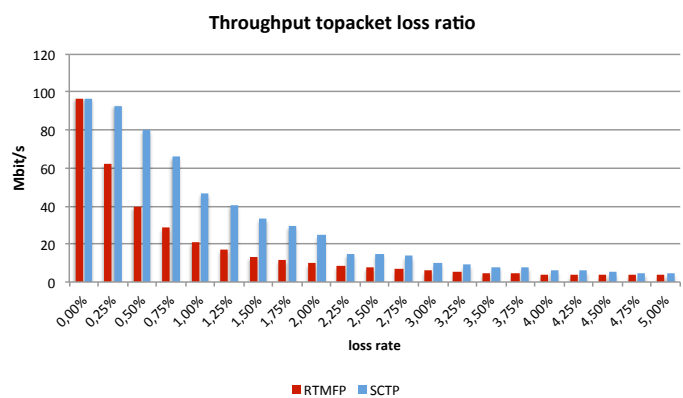

Fig. 5. Bottleneck scenario with two RTMFP peers and random traffic

## VI. Conclusion and outlook

The implemented RTMFP simulation model covers nearly all specified features of the protocol which are relevant for connection establishment and data transfer, i.e. four way handshake, congestion control, flow control, chunk bundling and support for multiple sessions and flows. This allows a detailed analysis of the built-in congestion control and flow control mechanisms.

With the new model we were able to compare RTMFP to the quite similar SCTP and measure the performance of both protocols under different conditions. We are currently working on a connecting module to use INET's Netperfmeter module with RTMFP. This will allow us to use the Netperfmeter as a traffic generator for RTMFP and SCTP. We are planning to contribute the RTMFP model to the INET suite.

## References

[1] M. Thornburgh, "Adobe's Secure Real-Time Media Flow Protocol", RFC 7016, November 2013

[2] S. Loreto, S.P. Romano, "Real-Time Communications in the Web: Issues, Achievements, and Ongoing Standardization Efforts", Internet Computing, IEEE (Volume:16 , Issue: 5 ) Page(s): 68 - 73 Sept.-Oct. 2012

[3] M. Thornburgh, "Adobe's RTMFP Profile for Flash Communication", RFC 7425, December 2014

[4] R. Stewart, Ed., "Stream Control Transmission Protocol", RFC 4960, September 2007

[5] R. Jesup, S. Loretto, M. Tüxen, "WebRTC Data Channels", draft-ietf-rtcweb-data-channel-13, January 4, 2015

[6] M. Tüxen, R. Stewart, R. Jesup, S. Loreto, "DTLS Encapsulation of SCTP Packets", draft-ietf-tsvwg-sctp-dtls-encaps-09, January 24, 2015

[7] I. Rüngeler, M. Tüxen, E. Rathgeb, "Integration of SCTP in the OMNeT++ Simulation Environment", Proc. of the 1st international conference on Simulation tools and techniques for communications, networks and systems workshops, March 3, 2008